\documentstyle[axodraw,epsfig,sprocl]{article}

\def\lsim{\raise0.3ex\hbox{$\;<$\kern-0.75em\raise-1.1ex\hbox{$\sim\;$}}}
\def\gsim{\raise0.3ex\hbox{$\;>$\kern-0.75em\raise-1.1ex\hbox{$\sim\;$}}}

\def\etal{{\it et al.}}
\def\half{{\textstyle{1 \over 2}}}

\def\bold#1{\setbox0=\hbox{$#1$}
     \kern-.025em\copy0\kern-\wd0
     \kern.05em\copy0\kern-\wd0
     \kern-.025em\raise.0433em\box0 }

\def\be{\begin{equation}}
\def\ee{\end{equation}}
\def\bea{\begin{eqnarray}}
\def\eea{\end{eqnarray}}

\begin{document}

\rightline{\small{hep-ph/9910445}}
\rightline{\small{UCCHEP/4-99}}
\rightline{\small{FSU-HEP-991020}}
\vskip+0.1cm

\title{ONE--LOOP RADIATIVE CORRECTIONS TO CHARGINO PAIR PRODUCTION}

\author{MARCO AURELIO D\'IAZ}

\address{\vskip 0.3cm Facultad de F\'\i sica, Universidad Cat\'olica de Chile \\
Av. Vicu\~na Mackena 4860, Santiago, Chile \\
and \\
Department of Physics, Florida State University \\ Tallahassee, Florida 
32306, USA}


\maketitle\abstracts{
We report on the effect that one--loop radiative corrections to the chargino 
pair production cross section has on the determination of the fundamental 
parameters of the theory. We work in the context of electron--positron 
colliders with $\sqrt{s}=500$ GeV. We conclude that the inclusion of these
corrections is crucial in precision measurements, specially at large and
small values of $\tan\beta$.
}

In the Minimal Supersymmetric Standard Model (MSSM) the supersymmetric 
fermionic partners of the $W$ gauge bosons and the charged Higgs $H^{\pm}$
mix to form a set of two Dirac fermions called charginos 
$\tilde\chi^{\pm}_i$, $i=1,2$. The chargino mass matrix is well known:
\begin{equation}
{\bf{\cal{M}}_C}=\left[\matrix{ M & \sqrt{2}m_W\cos\beta \cr
                                \sqrt{2}m_W\sin\beta & \mu }\right]
\label{chargmassmat}
\end{equation}
where $M$ is the gaugino mass associated to the $SU(2)$ group, $\mu$ is
the supersymmetric higgsino mass, and $\tan\beta=v_2/v_1$ is the ratio of 
the two Higgs vacuum expectation values (vev). This mass matrix is 
diagonalized by two rotation matrices $\bf U$ and $\bf V$ such that
\begin{equation}
{\bf U}^*{\bf{\cal{M}}_C}{\bf V}^{-1}={\mathrm{diag}}(m_{\tilde\chi^{\pm}_1},
m_{\tilde\chi^{\pm}_2})\,,
\label{diagmass}
\end{equation}
and every chargino interaction depends on the matrix elements $U_{ij}$ and 
$V_{ij}$.

Electron--Positron colliders are a specially clean environment for chargino
searches. They are produced in the s--channel via $Z$ and $\gamma$ gauge
bosons, and in the t--channel via electron--sneutrinos:

\begin{center}
\begin{picture}(160,50)(0,25) 
\ArrowLine(0,20)(20,40)
\ArrowLine(20,40)(0,60)
\Photon(20,40)(45,40){3}{6.5}
\ArrowLine(65,20)(45,40)
\ArrowLine(45,40)(65,60)
\Text(100,40)[]{$+$}
\ArrowLine(155,50)(135,60)
\ArrowLine(135,20)(155,30)
\DashLine(155,30)(155,50){3}
\ArrowLine(175,20)(155,30)
\ArrowLine(155,50)(175,60)
\Text(-8,60)[]{$e^+$}
\Text(-8,20)[]{$e^-$}
\Text(127,60)[]{$e^+$}
\Text(127,20)[]{$e^-$}
\Text(75,60)[]{$\tilde\chi^+_i$}
\Text(75,20)[]{$\tilde\chi^-_j$}
\Text(185,60)[]{$\tilde\chi^+_i$}
\Text(185,20)[]{$\tilde\chi^-_j$}
\Text(32,49)[]{$Z,\gamma$}
\Text(163,40)[]{$\tilde\nu_e$}
\end{picture}
\end{center}

\vskip +0.5cm\noindent
The measurement of the total production cross section, the chargino mass, and
the neutralino mass (a decay product), give enough information that
can be used to find the fundamental parameters of the theory \cite{DK-FS}.
This analysis also has been extended to CP violating scenarios and to
polarized beams \cite{chaparam}, and to the neutralino sector 
\cite{neutralinos}.

In order to have reliable results for the fundamental parameters of the 
theory extracted from chargino observables it is necessary to include the
one--loop radiative corrections to masses and cross section \cite{DKR,KNPY}.
Ref.~\cite{DKR} calculated these quantum corrections including the
leading Feynman graphs which contain quarks and squarks, since they are
enhanced by large Yukawa couplings. These kind of graphs correct vertices
and two point functions in such a way that the corrected amplitudes can
be represented by

\begin{center}
\begin{picture}(160,50)(0,25) 
\ArrowLine(0,20)(20,40)
\ArrowLine(20,40)(0,60)
\Photon(20,40)(45,40){3}{6.5}
\ArrowLine(65,20)(45,40)
\ArrowLine(45,40)(65,60)
\GCirc(47,40){5}{0.3}
\Text(100,40)[]{$+$}
\ArrowLine(155,50)(135,60)
\ArrowLine(135,20)(155,30)
\DashLine(155,30)(155,50){3}
\ArrowLine(175,20)(155,30)
\ArrowLine(155,50)(175,60)
\GCirc(155,30){5}{0.3}
\GCirc(155,50){5}{0.3}
\Text(-8,60)[]{$e^+$}
\Text(-8,20)[]{$e^-$}
\Text(127,60)[]{$e^+$}
\Text(127,20)[]{$e^-$}
\Text(75,60)[]{$\tilde\chi^+_i$}
\Text(75,20)[]{$\tilde\chi^-_j$}
\Text(185,60)[]{$\tilde\chi^+_i$}
\Text(185,20)[]{$\tilde\chi^-_j$}
\Text(32,49)[]{$Z,\gamma$}
\Text(163,40)[]{$\tilde\nu_e$}
\end{picture}
\end{center}

\vskip +0.5cm\noindent
with the renormalized vertices parametrized by form factors as defined in 
\cite{DKR}. Each sneutrino vertex has only one form factor:

\begin{center}
\begin{picture}(80,30)(22,28) 
\ArrowLine(35,20)(55,30)
\DashLine(55,30)(55,50){3}
\ArrowLine(75,20)(55,30)
\GCirc(55,30){5}{0.3}
\Text(27,20)[]{$e^-$}
\Text(85,20)[]{$\tilde\chi^-_j$}
\Text(63,48)[]{$\tilde\nu_e$}
\end{picture}
$
=iC^{-1}(1+\gamma_5)F^-_{\tilde\nu_e}\,,
$
\end{center}
\vskip+0.7cm
\noindent
where $C$ is the charge conjugation matrix. The form factor 
$F^-_{\tilde\nu_e}$ receive contributions from chargino mixing and chargino 
wave function renormalization. In order to be non--vanishing the charginos
have to be a mixing between higgsino and gaugino. In this case, corrections
proportional to logarithms of squark masses are enhanced by large Yukawa 
couplings.

The $Z$ gauge boson vertex has the following form factors:
\begin{center}
\begin{picture}(80,35)(8,38) 
\Photon(20,40)(45,40){3}{6.5}
\ArrowLine(65,20)(45,40)
\ArrowLine(45,40)(65,60)
\GCirc(47,40){5}{0.3}
\Text(75,60)[]{$\tilde\chi^+_i$}
\Text(75,20)[]{$\tilde\chi^-_j$}
\Text(25,49)[]{$Z$}
\end{picture}
$
=i\,{\cal{G}}^{ij}_{Z\chi\chi}
$
\end{center}
\vskip+0.9cm
\noindent
with
\begin{equation}
{\cal{G}}^{ij}_{Z\chi\chi}=(1+\gamma_5)
\big[F^+_{Z0}\gamma^{\mu}+F^+_{Z1}k_1^{\mu}+F^+_{Z2}k_2^{\mu}\big]
+(1-\gamma_5)
\big[F^-_{Z0}\gamma^{\mu}+F^-_{Z1}k_1^{\mu}+F^-_{Z2}k_2^{\mu}\big]
\end{equation}
and $k_1^{\mu}$ ($k_2^{\mu}$) is the 4--momenta of the chargino 
$\tilde\chi^-_j$ ($\tilde\chi^+_i$). Analogous expressions are valid for 
the photon form factors. In ${\cal{G}}_{Z\chi\chi}$ we include triangular
1PI graphs, chargino mixing and self energies, and gauge bosons mixing and 
self energies. The corrections turn out to be enhanced by large Yukawa
couplings $h_t$ and $h_b$, and proportional to logarithms of the squark 
masses. The corrections for this center of mass energy of 500 GeV can
have either sign and go up to values of $\pm20\%$, $\pm15\%$, or $\pm5\%$
for squark mass parameter given by 1 TeV, 600 GeV, or 200 GeV \cite{DKR}.

In the following four figures we plot the tree--level (dashes) and 
one--loop corrected (solid) total production cross section 
$\sigma(e^+e^-\rightarrow\tilde\chi^+_1\tilde\chi^-_1)$ as a function of
the neutralino mass (LSP). We keep constant the squark mass parameters
$M_Q=M_U=M_D=1$ TeV as well as the trilinear couplings $A_U=A_D=1$ TeV,
and we work with a center of mass energy $\sqrt{s}=500$ GeV.

\begin{figure}[htb]
\vspace{0.1cm}
\centerline{\protect\hbox{\psfig{file=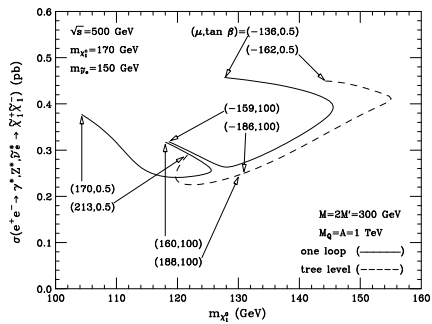,height=5.6cm,angle=90}}}
\vskip -0.1cm
\caption{Tree--level and one--loop corrected production cross section of
a pair of charginos as a function of the LSP mass. We take 
$m_{\tilde\chi^+_1}=170$ GeV, $m_{\tilde\nu_e}=150$ GeV, and $M=2M'=300$
GeV.}
\vskip -0.2cm
\label{c-n1_5a}
\end{figure}

In Fig.~\ref{c-n1_5a} we consider a constant, one--loop corrected, chargino
mass $m_{\tilde\chi^+_1}=170$ GeV, a sneutrino mass given by 
$m_{\tilde\nu_e}=150$ GeV, and gaugino masses $M=2M'=300$ GeV. The parameter
$\tan\beta$ is varied along the curve with extreme values 
$0.5<\tan\beta<100$ and, considering that the chargino mass is constant, this
fixes the value of $|\mu|$. Two branches appear according to the sign of
$\mu$. In parenthesis, and indicated by arrows it is shown the extreme
values of $\tan\beta$ and $\mu$. It is obvious from the figure that the
largest deviations occur for extreme values of $\tan\beta$. 

\begin{figure}[htb]
\vspace{-0.2cm}
\centerline{\protect\hbox{\psfig{file=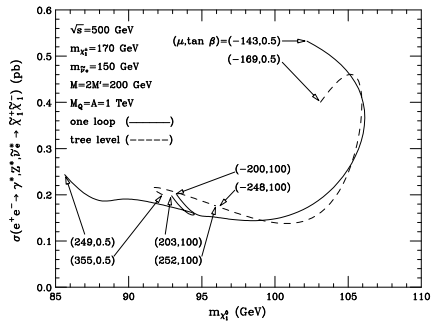,height=5.6cm,angle=90}}}
\vskip -0.1cm
\caption{Tree--level and one--loop corrected production cross section of
a pair of charginos as a function of the LSP mass. We take 
$m_{\tilde\chi^+_1}=170$ GeV, $m_{\tilde\nu_e}=150$ GeV, and $M=2M'=200$
GeV.}
\vskip -0.5cm
\label{c-n1_5b}
\end{figure}
In Fig.~\ref{c-n1_5b} we change the gaugino masses to $M=2M'=200$ keeping 
the previous chargino mass $m_{\tilde\chi^+_1}=170$ GeV and sneutrino mass
$m_{\tilde\nu_e}=150$ GeV. The main effect is to lower the possible values 
of the neutralino mass, which is bounded from above approximately by
$m_{\tilde\chi^0_1}\lsim\half M=M'$ (the LSP is mainly gaugino).
Notice the proximity of the tree level point $(\mu,\tan\beta)=(355,0.5)$
and the one--loop corrected $(203,100)$ which can give and idea of the 
confusion it may cause the non inclusion of radiative corrections.

\begin{figure}[htb]
\vspace{0.1cm}
\centerline{\protect\hbox{\psfig{file=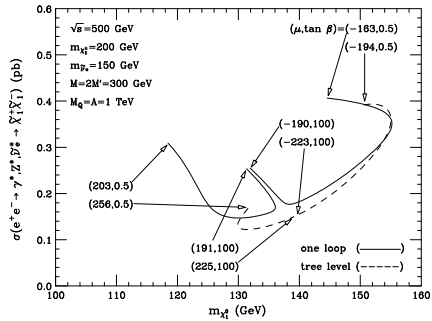,height=5.6cm,angle=90}}}
\vskip -0.1cm
\caption{Tree--level and one--loop corrected production cross section of
a pair of charginos as a function of the LSP mass. We take 
$m_{\tilde\chi^+_1}=200$ GeV, $m_{\tilde\nu_e}=150$ GeV, and $M=2M'=300$
GeV.}
\vskip -0.2cm
\label{c-n1_5c}
\end{figure}
In Fig.~\ref{c-n1_5c} we have increased the chargino mass with respect to the
first figure. Now we have $m_{\tilde\chi^+_1}=200$ GeV, $m_{\tilde\nu_e}=150$ 
GeV, and $M=2M'=300$ GeV. The values of $\mu$ are close to the chargino mass,
indicating that it is mainly higgsino. The total cross section is smaller,
and corrections to the tree level value can be as high as $50\%$.

\begin{figure}[htb]
\vspace{0.1cm}
\centerline{\protect\hbox{\psfig{file=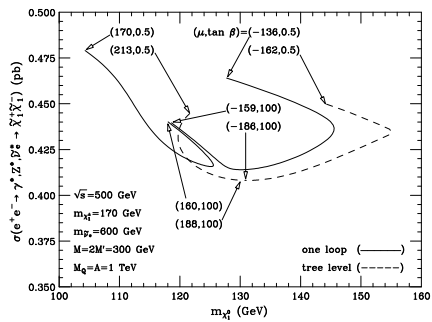,height=5.6cm,angle=90}}}
\vskip -0.1cm
\caption{Tree--level and one--loop corrected production cross section of
a pair of charginos as a function of the LSP mass. We take 
$m_{\tilde\chi^+_1}=170$ GeV, $m_{\tilde\nu_e}=600$ GeV, and $M=2M'=300$
GeV.}
\vskip -0.2cm
\label{c-n1_5d}
\end{figure}
Finally, in Fig.~\ref{c-n1_5d} we have change the sneutrino mass to
$m_{\tilde\nu_e}=600$ GeV compared with the first figure. We maintain the
value of the chargino mass $m_{\tilde\chi^+_1}=170$ GeV and the gaugino 
masses $M=2M'=300$ GeV. The effect of increasing the sneutrino mass is to
increase the total cross section where there was a large interference 
effect. In this case there is also a potential confusion in the region of
parameter space where the tree level curve at low values of $\tan\beta$ and
positive $\mu$ intersect the one--loop corrected curves at very large 
values of $\tan\beta$.

\begin{table}
\begin{center}
\caption{Tree--level and one--loop chargino observables from the second case 
study in ref.~[6].}
\bigskip
\begin{tabular}{lccccc}
\hline
observable & tree--level & one--loop & correction \\
\hline
$m_{\tilde\chi^+_1}$ (GeV) & 109.9   & 112.6  & $+2.5\%$ \\
$m_{\tilde\chi^+_2}$ (GeV) & 342.3   & 359.0  & $+4.8\%$ \\
$\sigma_{11}$ \,\,\,(pb)   & 0.73    & 0.76   & $+4.3\%$ \\
$\sigma_{12}$ \,\,\,(fb)   & 8.4     & 7.0    & $-16\%$  \\
\hline
\label{tab:cases}
\end{tabular}
\vskip-0.7cm
\end{center}
\end{table}
As an example, we consider the second case study in ref. \cite{BDFT}, 
motivated by $SO(10)$ models with non--universal boundary conditions at 
the GUT scale due to extra D--terms contributions. In this case we have
$M=116.4$ GeV, $\mu=-320.8$ GeV, $m_{\tilde\nu_e}=1018.2$ GeV, and 
$\tan\beta=47$, typical of $SO(10)$ models with top--bottom--tau Yukawa
unification. Squark masses are of the order of 800 GeV. We observe a 
$4.3\%$ correction to the total cross section 
$\sigma_{11}=\sigma(e^+e^-\rightarrow\tilde\chi^+_1\tilde\chi^-_1)$,
and comparable corrections for the chargino masses. The corrections on 
$\sigma_{12}=\sigma(e^+e^-\rightarrow\tilde\chi^+_1\tilde\chi^-_2)+
\sigma(e^+e^-\rightarrow\tilde\chi^+_2\tilde\chi^-_1)$ is larger and 
negative ($-16\%$). As a point of comparison we mention that it is 
expected to measure experimentally the chargino mass at the 0.1\% level.
We conclude that the inclusion of one--loop radiative corrections to the 
chargino observables is necessary in order to obtain reliable results in
the determination of the fundamental parameters from experimental 
measurements.

\section*{Acknowledgments}

I am thankful to my collaborators S.F. King and D.A. Ross for their 
invaluable contribution to the work presented here. I also thank H. Baer 
and the Physics Department of the Florida State University for their 
hospitality, where I was supported by the U.S. DOE contract number 
DE-FG02-97ER41022.

\end{document}